\documentclass[onecolumn, floatfix, preprintnumbers, eqsecnum, letterpaper, superscriptaddress, nofootinbib]{revtex4}
\usepackage{graphicx}
\usepackage{microtype}
\usepackage{amsmath}
\usepackage{amssymb}
\usepackage{subfigure}
\usepackage{hyperref}
\usepackage{url}
\usepackage{xcolor}
\usepackage{color}
\usepackage{mathrsfs}
\usepackage{calrsfs}
\usepackage{amsfonts}
\usepackage{eufrak}
\usepackage{tabularx}
\usepackage{eucal}
\usepackage{latexsym}
\usepackage{ragged2e}
\usepackage{epsfig}
\usepackage{textcomp}

\usepackage{caption}
\usepackage{subfigure}

\usepackage{marvosym}
\usepackage{ifsym}

\usepackage{lipsum}

\DeclareCaptionJustification{justified}{\leftskip=0pt \rightskip=0pt \parfillskip=0pt plus 1fil}
\definecolor{vividviolet}{rgb}{0.62, 0.0, 1.0}
\definecolor{amaranth}{rgb}{0.9, 0.17, 0.31}
\definecolor{palatinateblue}{rgb}{0.15, 0.23, 0.89}
\definecolor{brightpink}{rgb}{1.0, 0.0, 0.5}
\definecolor{cornflowerblue}{rgb}{0.39, 0.58, 0.93}
\definecolor{deepcarminepink}{rgb}{0.94, 0.19, 0.22}
\definecolor{radicalred}{rgb}{1.0, 0.21, 0.37}

\hypersetup{ linktoc=all,
    colorlinks, linkcolor={palatinateblue},
    citecolor={brightpink}, urlcolor={amaranth}
}

\graphicspath{{Images/}}

\graphicspath{{Images/}}

\def\@fnsymbol#1{\ensuremath{\ifcase#1\or \ddagger \or  $\textleaf$  \or \dagger
\else\@ctrerr\fi}}%

\makeatother
\def\sideremark#1{\ifvmode\leavevmode\fi\vadjust{\vbox to0pt{\vss% the remark
 \hbox to 0pt{\hskip\hsize\hskip1em%                          will appear only
 \vbox{\hsize1.3cm\tiny\raggedright\pretolerance10000%          on the side
 \noindent #1\hfill}\hss}\vbox to8pt{\vfil}\vss}}}%

\def\beq{\begin{equation}}
\def\eeq{\end{equation}}

 \newcommand{\be}{\begin{equation}}
	\newcommand{\en}{\end{equation}}

\begin{document}

\title{Thermodynamics of the arbitrary dimensional FRW universe: Joule-Thomson expansion}

\author{Haximjan Abdusattar \Letter}
\email{axim@nuaa.edu.cn}
\affiliation{School of Physics and Electrical Engineering, Kashi University, Kashi 844006, Xinjiang, China}

\author{Shi-Bei Kong}
\email{shibeikong@ecut.edu.cn}
\affiliation{School of Science, East China University of Technology, \\
Nanchang 330013, Jiangxi, China}

\author{Minawar Omar}
\email{918932401@qq.com}
\affiliation{School of Physics and Electrical Engineering, Kashi University, Kashi 844006, Xinjiang, China}

\author{Zhong-Wen Feng}
\email{zwfengphy@cwnu.edu.cn}
\affiliation{School of Physics and Astronomy, China West Normal University,\\ Nanchong 637009, China}

{\let\thefootnote\relax\footnotetext{\vspace*{0.1cm}$^{\text{\Letter}}$ Corresponding Author}}

\begin{abstract}

In this paper, we investigate the thermodynamics especially the Joule-Thomson expansion of the $n$-dimensional FRW (Friedmann-Robertson-Walker) universe with a perfect fluid. We derive the thermodynamic equations of state $P=P(V, T)$ for the $n$-dimensional FRW universe in  Einstein gravity and Einstein-Gauss-Bonnet gravity, where the thermodynamic pressure $P$ is defined by the work density $W$ of the perfect fluid, $i.e.$ $P\equiv W$. Furthermore, we present the Joule-Thomson expansion as an application of these equations of state to elucidate the cooling-heating properties of the $n$-dimensional FRW universe. We determine the inversion temperature and inversion pressure in the FRW universe with arbitrary dimensions for the first time, and illustrate the characteristics of inversion curves and isenthalpic curves in the $T$-$P$ plane. We also examine constraints on the perfect fluid in the FRW universe, as derived from the cooling-heating transition point.
This study offers insights into deepening our comprehension of cooling and heating regions in the FRW universe, thereby revealing its expansion mechanisms.

\end{abstract}

\maketitle

\section{Introduction}

Since the discovery of Hawking radiation \cite{Hawking:1974sw} and the first proposal of Bekenstein-Hawking entropy \cite{Bardeen1973,Bekenstein:1974ax}, people have found that black holes can be viewed as thermodynamic systems, exhibiting interesting thermodynamic similarities to classical thermodynamic systems. Subsequent studies generalized this entropy concept to incorporate quantum corrections, gravitational effects in non-static spacetimes, and even cosmological horizons, leading to the framework of generalized entropy \cite{Bousso:1999xy,Wall:2010jtc} -- a key step linking black hole thermodynamics to broader gravitational systems. In the presence of a cosmological constant, Hawking and Page \cite{Hawking:1982dh} investigated the thermodynamics of Schwarzschild-AdS (anti-de Sitter) black holes and found an interesting phase transition between the black hole and thermal gas in AdS space. Since then, significant attention has been devoted to the study of asymptotically AdS black holes because they admit a gauge-gravity duality description \cite{Witten:1998zw}. This duality has not only deepened the understanding of black hole thermodynamics but also enabled its extension to holographic cosmology, where the entropy and thermodynamics of cosmological spacetimes ($e.g.,$ FRW universe) are holographically encoded in boundary quantum field theories \cite{Maldacena:1997re,Banks:2001px,Cai:2005ra}. In previous studies, the cosmological constant was treated as a fixed parameter. Studying the thermodynamic properties of black holes in AdS space is of great importance for understanding the AdS/CFT correspondence. In recent years, researchers have treated the negative cosmological constant $\Lambda$ as the thermodynamic pressure $P=-\Lambda/(8\pi)$ \cite{Kastor:2009wy,Dolan:2010ha} and discovered many other interesting thermodynamic phenomena in the extended phase space of AdS black holes, such as $P$-$V$ criticality \cite{Kubiznak:2012wp,Cai:2013qga,Xu:2015rfa,Hu:2018qsy,Abdusattar:2023xxs,Abdusattar:2024zzi}, Joule-Thomson expansion \cite{Okcu:2016tgt,Feng:2020swq,Barrientos:2022uit}, etc.

An interesting question is that whether the thermodynamic properties of the FRW universe share some similarities with black holes. A remarkable feature of thermodynamics is its universality. The largest known system controlled by gravity, namely the universe, should also obey the laws of thermodynamics.
According to the spirit of Jacobson \cite{Jacobson:1995ab} that the Einstein field equation can be derived from the Clausius relation, Cai and Kim \cite{Cai:2005ra} discovered that Hawking radiation also exist in the FRW universe, and studied the Friedmann equation of the FRW universe. Later, the unified first law of the FRW universe was also studied \cite{Cai:2006rs}, which after being projected onto the apparent horizon leads to the first law of thermodynamics. It has been shown that the Clausius relation holds for the FRW universe in Einstein gravity and some modified theories of gravity \cite{Akbar:2006kj,Akbar:2006er} - a result that also suggests the FRW universe is an (quasi-)equilibrium thermodynamic system \cite{Cai:2006pa,Wu:2009wp}.

As an extension, we have recently conducted first principle study of thermodynamic pressure for the four-dimensional FRW universe thermodynamics \cite{Abdusattar:2021wfv}.
We investigated the thermodynamics for the FRW universe, especially presented the thermodynamic equation of state for the four dimensional FRW universe in Einstein gravity, $i.e.$ $P = P(V, T)$ for the first time, where the thermodynamic pressure $P$ is defined by the work density $W$ of the perfect fluid \cite{Hayward:1993wb,Hayward:1994bu}
\begin{eqnarray}\label{WD}
P\equiv W:=-\frac{1}{2}h_{ij}T^{ij}\,,
\end{eqnarray}
here $h_{ij}$ and $T^{ij}$ are the $0,1$-components of the metric and the stress-tensor \cite{Akbar:2006kj} with $i,j=0,1, x^0=t, x^1=r$.
In addition, we extended the thermodynamic properties of the FRW universe to modified theories of gravity and found that the $P$-$V $phase transitions \cite {Kong:2021qiu,Abdusattar:2023hlj}. It is worth noting that the coexisting phase of the $P$-$V$ phase transition in the FRW universe is higher than the critical temperature, which distinguishes it from van der Waals systems and AdS black holes system.
It enables us to study the micro-structure of the FRW universe as a thermodynamic system \cite{Abdusattar:2023pck}. Our analysis of the Ruppeiner geometry for the FRW universe provided new insights into the nature of interactions among the perfect fluid matter constituents in the expanding FRW universe from the viewpoint of geometro-thermodynamics.
Subsequently, encouraged by the methods in the above works, the thermodynamic properties of the FRW universe with same method has also been studied in some of modified theories of gravity
\cite{Saavedra:2023lds,Housset:2023jcm}.

In this study, we aim to explore the thermodynamic properties of the arbitrary $n$-dimensional FRW universe within both Einstein gravity and Einstein-Gauss-Bonnet gravity frameworks.
We initiate our investigation by examining the thermodynamics of the $n$-dimensional FRW universe filled with a perfect fluid within the context of Einstein gravity, deriving the equation of state. Subsequently, we delve deeper into the thermodynamic properties of this universe, meticulously calculating its heat capacity and comparing it with the established results in the four-dimensional realm. Furthermore, we extend our exploration to embrace Einstein-Gauss-Bonnet gravity, a natural extension of Einstein's theory of general relativity motivated by profound considerations from string theory. Within this modified gravity framework, the Gauss-Bonnet term assumes significant importance in higher dimensions, potentially influencing the evolution and thermodynamic properties of the FRW universe, thus providing valuable insights into the thermodynamics of the universe.

Furthermore, as an application of thermodynamic equations of state, we study the Joule-Thomson expansion for the $n$-dimensional FRW universe.
In classical thermodynamics, the Joule-Thomson expansion process of a van der Waals system describes a gas moving from high pressure region to low pressure region at a fixed enthalpy. In \cite{Okcu:2016tgt}, the author creatively applied the famous Joule-Thomson expansion to the charged AdS black hole and found the existence of inversion temperature and inversion pressure. Subsequently, various black holes were studied, such as Gauss-Bonnet black holes \cite{Lan:2018nnp}, regular (Bardeen)-AdS black hole \cite{Pu:2019bxf}, Born-Infeld AdS black holes \cite{Bi:2020vcg}, hairy black holes \cite{Xing:2021gpn}, nonlinearly charged AdS black holes \cite{Kruglov:2022lnc}, see  \cite{Ghaffarnejad:2020nhb,Nam:2019zyk,Dehghani:2023yph,Du:2023kgj,Feng:2022wnq} for more related works.
In our previous work, we have studied the Joule-Thomson expansion of the four-dimensional FRW universe under Einstein gravity \cite{Abdusattar:2021wfv} and braneworld scenario \cite{Kong:2022xny}\footnote{In the braneworld scenario, our universe is modeled as a brane with three spatial dimensions embedded in a higher dimensional bulk, see for \cite{Meehan:2014zsa,Abdusattar:2015azp} related works and the references therein.}, hence found the existence of inversion temperature and inversion pressure.
As a generalization, we extend the research on the Joule-Thomson expansion of the FRW universe to the Einstein gravity and Einstein-Gauss-Bonnet gravity models with arbitrary dimensions. We introduce the $T$-$P$ isenthalpic expansion process in a high-dimensional FRW universe, pioneering the exploration of thermodynamic properties in arbitrary dimensions for the first time.
Drawing upon these investigations, we aim to deduce the thermodynamic characteristics of the higher-dimensional FRW universe and further assess how Einstein-Gauss-Bonnet gravity modifies these properties, offering a comprehensive understanding of the thermodynamic behavior in such systems.

The organization of this paper is as follows. In Sec.\ref{secII}, we study the thermodynamics of the $n$-dimensional FRW universe, and derive the equation of state $P = P(V, T)$ in Einstein gravity.
In Sec.\ref{secIII}, we obtain the equation of state $P = P(V, T)$ for the $n$-dimensional FRW universe in a Einstein-Gauss-Bonnet gravity, which is a natural and most effective generalization of Einstein gravity.
In Sec.\ref{secIV}, we study the Joule-Thomson expansion for the $n$-dimensional FRW universe as an application of the thermodynamic equations of state.
In Sec.\ref{secV}, we give the conclusions and discussions.
Throughout this paper, we use natural units such that $c=\hbar=k_{B}=1$.

\section{Einstein Gravity: Thermodynamics and Equation of State of Arbitrary Dimensional FRW Universe}\label{secII}

In this section, we review the apparent horizon and Hawking temperature of the FRW universe. Then we study the thermodynamics of the $n$-dimensional FRW universe in Einstein gravity, and construct the equation of state $P=P(V, T)$ in which the thermodynamic pressure $P$ is defined by the work density $W$ of the perfect fluid.

\subsection{Apparent Horizon and Hawking Temperature}

We consider a spatially homogeneous and isotropic universe described by the FRW metric, where the line-element of an arbitrary $n$-dimensional FRW universe is expressed as
\begin{equation}\label{FRWds}
 ds^2= -dt^2 + a^2(t) \Big[\frac{dr^2}{1-kr^2} + r^2 d \Omega_{(n-2)}^2\Big]\,,
\end{equation}
where $a(t)$ is the time-dependent scale factor,
$k=-1$, $+1$, $0$ is the spatial curvature corresponding to an open (hyperbolic), closed (spherical), and flat universe, and $d \Omega_{(n-2)}^2$ represents the metric of the $(n-2)$-dimensional unit sphere.

For convenience of later discussion, we rewrite the line-element of $n$-dimensional FRW universe (\ref{FRWds}) with physical radius $R\equiv a(t)r$ in the following form
\begin{equation}\label{NewMetric}
ds^{2}=h_{ij}dx^i dx^j+R^2d \Omega_{(n-2)}^2\,,
\end{equation}
where $i,j=0,1$ with $x^0=t, x^1=r$ and $h_{ij}={\rm diag} [-1, a^2(t)/(1-kr^2)]$. For simplicity, we denote $a \equiv a(t)$ in the following. For the dynamical spacetime, one can define an apparent horizon by the marginally trapped surface with vanishing expansion, $i.e.$ $(D R)^2:=h^{ij}(D_i R)(D_j R)=0$ \cite{Hayward:1993wb}. Apply this condition to the metric (\ref{NewMetric}), the apparent horizon radius of the FRW universe is obtained \cite{Cai:2005ra}
\begin{equation}\label{RA}
 R_A = \frac{1}{\sqrt{H^2 +k/a^2}}\,,
 \end{equation}
whose time derivative is
 \begin{equation}\label{dRA}
 \dot{R}_A= -HR_A^3 \Big(\dot H-\frac{k}{a^2}\Big)\,,
 \end{equation}
where $H\equiv\dot{a}/a$ is the Hubble parameter, and $``\cdot"$ stands for the derivative with respect to the cosmic time.
The above equation (\ref{dRA}) shows the time dependence of the apparent horizon.

The surface gravity at the apparent horizon of the FRW universe is given by \cite{Cai:2005ra}
\begin{equation}\label{SurfaceG}
\kappa=-\frac{1}{R_A}\Big(1-\frac{\dot{R}_A}{2HR_A}\Big)\,.
\end{equation}
%which is negative
%\footnote{This means that the apparent horizon of the FRW universe is an inner trapping horizon \cite{Hayward:1993wb}.}
%in this paper because we treat $\dot{R}_A$ as a small quantity, i.e. $\dot{R}_A<2HR_A$.
Assuming $\dot{R}_A$ to be a small quantity, the surface gravity $\kappa$ of the apparent horizon in the FRW universe is negative\footnote{See relevant previously works on negative surface gravity in Refs. \cite{Abdusattar:2021wfv,Hayward:1993wb,Abdusattar:2022bpg,Dolan:2013ft}.}.
The Hawking temperature of the FRW universe is defined from the surface gravity (\ref{SurfaceG})
\begin{alignat}{1}
T\equiv\frac{|\kappa|}{2\pi}=\frac{1}{2\pi R_A}\Big(1-\frac{\dot{R}_A}{2HR_A}\Big)\,. \label{HawkingT}
\end{alignat}
Here, we assume that $\dot{R}_A \ll 1$ and concentrate on the situation of approximate thermal equilibrium, where the FRW universe undergoes an extremely gradual evolution, resulting in all thermodynamic processes being quasistatic.

\subsection{The Equation of State of Arbitrary Dimensional FRW Universe}

Now, we study the thermodynamics especially the equation of state for the $n$-dimensional $(n\geq 4)$ FRW universe in Einstein gravity.
The action in the $n$-dimensional $(n\geq 4)$ spacetime with matter fields is given by
\begin{equation}\label{action}
S=\frac{1}{2k_n^2}\int d^{n} x \sqrt{-g} \mathcal{R} + S_{m}\,,
\end{equation}
where $\mathcal{R}$ is the $n$--dimensional Ricci scalar, $k_n \equiv \sqrt{8 \pi G_{n}}$, in which $G_{n}$ is the $n$-dimensional gravitational constant, $S_{m}$ is the action of the matter fields. From Eq. (\ref{action}), the Einstein field equation is given by
\begin{equation}\label{FieldEq}
R_{\mu\nu}-\frac{1}{2}g_{\mu\nu}\mathcal{R}=k_n^2 T_{\mu\nu}\,,
\end{equation}
where $T_{\mu\nu}$ is the energy momentum tensor of the matter field. In the FRW universe, we assume that the energy momentum tensor is described by perfect (isotropic) fluid
\begin{equation}\label{Tmunu}
T_{\mu\nu}=(\rho+p)u_{\mu}u_{\nu}+p g_{\mu\nu}\,,
\end{equation}
where $\rho$ and $p$ are the energy density and pressure, and $u_{\mu}$ is the four-velocity of the perfect fluid satisfying $u_\mu u^\mu=-1$.

Applying the field equations (\ref{FieldEq}) to the line-element (\ref{FRWds}) with the Eq.(\ref{Tmunu}) one gets the Friedmann's equations
\begin{eqnarray}\label{FE}
H^{2}+\frac{k}{a^{2}}=\frac{16 \pi G_{n}}{(n-1)(n-2)} \rho \,, \quad \dot{H}-\frac{k}{a^2}=-\frac{8 \pi G_{n}}{n-2}(\rho+p)\,.
\end{eqnarray}
From Eqs.(\ref{RA}), (\ref{dRA}) and (\ref{FE}), one can express $\rho$ and $p$ in terms of $R_A$ and $\dot{R}_A$, $i.e.$
\begin{eqnarray}
\rho&=&\frac{(n-1)(n-2)}{16\pi G_{n} R_A^2}\,,\label{rho1}\\
p&=&-\frac{(n-1)(n-2)}{16\pi G_{n} R_A^2}+\frac{(n-2)}{8\pi G_{n}}\frac{\dot{R}_A}{H R_A^3}\,.\label{pm1}
\end{eqnarray}

From Eq.(\ref{WD}) with (\ref{rho1}) and (\ref{pm1}), we obtain the work density of the $n$-dimensional FRW universe associated with perfect fluid
\begin{alignat}{1}
W=\frac{1}{2}(\rho-p)=\frac{(n-1)(n-2)}{16\pi G_{n} R^2_A}-\frac{(n-2)\dot{R}_A}{16\pi G_{n} H R^3_A}\,.  \label{WD1}
\end{alignat}

The Misner-Sharp energy within the framework of Einstein gravity for the $n$-dimensional FRW universe inside the apparent horizon is given by \cite{Maeda:2007uu,Cai:2009qf}
\begin{equation}\label{E}
 E=\frac{(n-2)V_{n-2}^{1} {R^{n-3}_{A}}}{16\pi G_{n}}\,.
\end{equation}
Differentiating the above equation, we easily obtain the following relation
%\footnote{Note that the minus sign before $T dS$ in (\ref{FL}) is caused by the negative surface gravity, $i.e.$ $\kappa<0$, so keeping the corresponding temperature $T$ positive will impose constraints on the perfect fluid. The relevant derivation is shown in the Appendix (\ref{AnkappaFRW}).}
\begin{equation}
d E=-Td S+Wd V \,, \label{FL}
\end{equation}
where $S$ is the Bekenstein-Hawking entropy \cite{Bekenstein:1973ur,Abdusattar:2023fdm}
\begin{equation}\label{S}
S= \frac{A}{4 G_{n}} \,
%=\pi^{(n-1)/2}R_A^{n-2} / 2\Gamma[{(n-1)}/{2}],
\end{equation}
with $A:=V_{n-2}^{1} R_A^{n-2}$ is the area associated with apparent horizon, where $V_{n-2}^{1}$ is the surface area of the ($n-2$)-dimensional unit Einstein space defined by $ V_{n-2}^{1}:=2\pi^{(n-1)/2} / \Gamma[{(n-1)}/{2}]$, and $V$ is the thermodynamic volume
\begin{equation}\label{TV}
V:=\frac{V_{n-2}^{1}}{n-1} R_A^{n-1}\,,
\end{equation}
which is the conjugate variable of the work density $W$.

Compare the Eq.(\ref{FL}) with the standard form of thermodynamic first law \cite{Hayward:1997jp,Wu:2009wp}
\begin{alignat}{1}
d U=Td S-Pd V\,, \label{SFL}
\end{alignat}
we see that the internal energy $U$ and thermodynamic pressure $P$ can be identified with $-E$ and $W$, $i.e.$
\begin{alignat}{1}
U:=&-E\,,
\\
P:=&W\,. \label{tp}
\end{alignat}
%Note that this definition of $P$ using the work density is more natural than that in the literature using the cosmological constant, in the sense that it is here a true variable rather than a constant.

The thermodynamic equation of state for the $n$-dimensional FRW universe in Einstein gravity from Eqs.(\ref{tp}), (\ref{HawkingT}) and (\ref{WD1}) is obtained by\footnote{From now on, we set $G_n = 1$ for simplicity in this section.}
\begin{alignat}{1}
P=\frac{(n-2)}{4} \frac{T}{R_A}+\frac{(n-2)(n-3)}{16\pi}\frac{1}{R^2_A}\,.\label{ES}
\end{alignat}
The essential question is whether this system undergoes a $P$-$V$ phase transition, which necessitates that the equation \cite{Kubiznak:2012wp,Hu:2018qsy,Kong:2021qiu,Abdusattar:2023hlj}
\begin{alignat}{1}
\Big(\frac{\partial P}{\partial V}\Big)_{T}=\Big(\frac{\partial^2 P}{\partial V^2}\Big)_{T}=0\,,\label{PV}
\end{alignat}
or equivalently
\begin{alignat}{1}
\Big(\frac{\partial P}{\partial R_A}\Big)_{T}=\Big(\frac{\partial^2 P}{\partial R^2_A}\Big)_{T}=0\,\label{PVTc}
\end{alignat}
has a critical-point solution $T=T_c,\ P=P_c,\ R_A=R_c$.
By substituting (\ref{ES}) into (\ref{PVTc}), one can easily check that no such solution exists, and thus there is no
$P$-$V$ phase transition in the arbitrary dimensional FRW universe with a perfect fluid in Einstein gravity.

Thermodynamic stability means a system's ability to stay in or return to a balanced state when small changes occur. For instance, the heat capacity at constant pressure $C_{P}$ is a key indicator: a positive $C_{P}$ corresponds to a stable system, while a negative $C_{P}$ indicates an unstable one. By using Eqs. (\ref{S}) and (\ref{ES}), we can derive the heat capacity of the $n$-dimensional FRW universe within the framework of Einstein gravity
\begin{eqnarray}\label{CP0}
C_{P}=\Big(\frac{\partial {\cal H}}{\partial T}\Big)_{P}=T\Big(\frac{\partial S}{\partial T}\Big)_{P}=
\frac{(n-2)\pi^{(1+n)/2} R_A^{n-1}T}{\Gamma[(n-1)/2](n-3+2\pi R_A T)}\,,
\end{eqnarray}
where ${\cal H}\equiv U+P V$ is the enthalpy of the system, and its specific expression is given in Appendix {(\ref{HHE})}.
Clearly, the heat capacity $C_P$ is always positive, which indicates that the universe governed by Einstein gravity is thermodynamically stable. It is worth noting that when $n = 4$, the value of $C_P$ is consistent with the result obtained in our previous work \cite{Abdusattar:2021wfv}, and this consistency also reflects the stability. This thus demonstrates that the thermodynamic stability of the FRW universe under Einstein gravity holds steady, irrespective of variations in spacetime dimension.

\section{Einstein-Gauss-Bonnet Gravity: Thermodynamics and Equation of State for Arbitrary Dimensional FRW Universe}\label{secIII}

Einstein-Gauss-Bonnet gravity is an extension of Einstein's theory of gravity that incorporates the Gauss-Bonnet term, a topological invariant that does not contribute in four-dimensional spacetime but gains significance in higher dimensions.
In this section, we delve into the thermodynamics of the $n$-dimensional FRW universe within the framework of Einstein-Gauss-Bonnet gravity, aiming to derive the thermodynamic equation of state.

\subsection{A Brief Introduction of Einstein-Gauss-Bonnet Gravity and Friedmann's Equation}

In this part, we make a brief introduction of Einstein-Gauss-Bonnet gravity and Friedmann's equation for the $n$-dimensional FRW universe. We commence with a concise introduction to Einstein-Gauss-Bonnet gravity, whose action in an $n(\geq 5)$-dimensional spacetime is given by
\cite{Maeda:2007uu,Hu:2010sn,Wu:2021zyl}
\begin{equation}\label{fullaction}
S_{GB}=\frac{1}{2k_n^2} \int d^{n} x \sqrt{-g}(\mathcal{R}+\alpha L_{\text{GB}}) + S_{m}\,,
\end{equation}
where $\alpha$ is a Gauss-Bonnet coupling constant\footnote{In string theory, the coupling constant $\alpha$ is typically positive, reflecting the inverse string tension \citep{Gross:1986iv,Gross:1986mw}. However, in our present work, we consider both positive and negative values of $\alpha$ for a more comprehensive analysis, aiming to explore novel physical phenomena beyond the standard framework.} with unit of length squared, and the Gauss-Bonnet term $L_{GB}$ expressed as
\begin{equation}\label{GBinvariant}
L_{\text{GB}}=\mathcal{R}^2-4 R^{\mu \nu} R_{\mu \nu}+R^{\mu \nu \sigma \gamma} R_{\mu \nu \sigma \gamma}\,,
\end{equation}
possesses topological invariance within the realm of four-dimensional spacetimes. Notably, Gauss-Bonnet gravity gracefully emerges as the second-order term within Lovelock gravity, maintaining the esteemed property that its equations of motion exclusively involve second-order derivatives of the metric. This elegant feature parallels the simplicity and elegance observed in Einstein gravity \citep{Lovelock:1971yv}.

By varying the equation (\ref{fullaction}) with respect to the metric tensor, one can derive the equation of motion \cite{Maeda:2007uu,Hu:2010sn,Kumar:2023cmo}
\begin{equation}\label{fieldequations}
R_{\mu\nu}-\frac{1}{2}g_{\mu\nu}\mathcal{R}+\alpha H_{\mu\nu}=k_n^2 T_{\mu\nu}\,,
\end{equation}
where
\begin{align} \label{GBcontribution}
    H_{\mu \nu}&=2\left(\mathcal{R} R_{\mu \nu}-2 R_{\mu \lambda} R_\nu^\lambda-2 R^{\gamma \delta} R_{\gamma \mu \delta \nu}+R_\mu^{\alpha \gamma \delta} R_{\alpha \nu \gamma \delta}\right) -\frac{1}{2} g_{\mu \nu} L_{G B}\nonumber
\end{align}
being the contribution coming from the Gauss-Bonnet term (\ref{GBinvariant}). As we know, for $n=4$, the tensor $H_{\mu\nu}$ is zero because the Gauss-Bonnet term (\ref{GBinvariant}) transforms into a topological invariant, which does not affect the dynamics of the theory.

Applying the modified theory of gravity (\ref{fieldequations}) to the FRW universe line element (\ref{FRWds}) with perfect fluid energy momentum stress-tensor (\ref{Tmunu}), the Friedmann's equations \cite{Cai:2005ra,Akbar:2006kj} are obtained as
\begin{eqnarray}\label{FE10}
\Big[1+\widetilde{\alpha}\Big(H^{2}+\frac{k}{a^{2}}\Big)\Big]\Big[H^{2}+\frac{k}{a^{2}}\Big]=\frac{16 \pi G_{n} \rho}{(n-1)(n-2)}  \,,
\end{eqnarray}
and
\begin{eqnarray}\label{FE11}
 \Big[1+2 \widetilde{\alpha}\Big(H^{2}+\frac{k}{a^{2}}\Big)\Big]\Big[\dot{H}-\frac{k}{a^2}\Big]=-\frac{8 \pi G_{n}}{n-2}(\rho+p) \,,
\end{eqnarray}
%which are also satisfy the energy conservation equation
%\begin{eqnarray}\label{continutyEq}
%\dot \rho + n H(\rho +p) =0\,,
%\end{eqnarray}
which are modified by the presence of the Gauss-Bonnet term via
$\widetilde{\alpha}\equiv (n-3)(n-4)\alpha$
 and is influenced by the spacetime dimension $n$ \citep{Cai:2002bn}. Notably, when $n=4$, the standard Friedmann's equations of the Einstein gravity scenario are recovered, as given in (\ref{FE}).

\subsection{The Equation of State of Arbitrary Dimensional FRW Universe}

In this part, within the framework of Einstein-Gauss-Bonnet gravity we obtain the thermodynamic equation of state for the $n$-dimensional FRW universe.

With the expressions of the apparent horizon (\ref{RA}) and (\ref{dRA}), the Friedmann's equations (\ref{FE10}) and (\ref{FE11}) can be rewritten as
\begin{eqnarray}
\rho&=&\frac{(n-1)(n-2)}{16\pi G_{n}R_A^2}\Big(1+\frac{\widetilde\alpha}{R_A^2}\Big)\,,\label{rho1}\\
p&=&-\frac{(n-1)(n-2)}{16\pi G_{n}R_A^2}\Big(1+\frac{\widetilde\alpha}{R_A^2}\Big)+\frac{(n-2)}{8\pi G_{n}}\Big(1+\frac{2\widetilde \alpha}{R_A^2}\Big)\frac{\dot{R}_A}{H R_A^3}\,.\label{pm1}
\end{eqnarray}
Substituting the above expressions into the definition of work density (\ref{WD}), we obtain the work density of the perfect fluid for a FRW universe in this modified gravity
\begin{eqnarray}\label{WDa}
W=\frac{1}{2}(\rho-p)=\frac{(n-1)(n-2)}{16\pi G_{n} R^2_A} \Big(1+\frac{\widetilde\alpha}{R^2_A}\Big)-\frac{(n-2)\dot{R}_A}{16\pi G_{n} H R^3_A} \Big(1+\frac{2\widetilde\alpha}{R^2_A}\Big) \,,
\end{eqnarray}
which coincides with the result discussed in \cite{Saavedra:2023lds}.

The Misner-Sharp energy evaluated at the FRW metric (\ref{NewMetric}) is given by \cite{Maeda:2007uu,Cai:2009qf}
\begin{equation}\label{EE}
    E=\frac{(n-2)V_{n-2}^{1} {R^{n-3}_{A}}}{16\pi G_{n}}\Big(1+\frac{\widetilde{\alpha}}{R^{2}_{A}}\Big)\,.
\end{equation}
Differentiating the above Misner-Sharp energy will actually lead to a good thermodynamic first law to the FRW universe in this Einstein-Gauss-Bonnet theory as
\begin{eqnarray}\label{dEE}
dE=-TdS+WdV \,,
\end{eqnarray}
where $V$ is the conjugate thermodynamic volume of $W$ and can still take the form of (\ref{TV}), $T$ and $W$ are given in (\ref{HawkingT}) and (\ref{WDa}), respectively. Note that the minus sign before $TdS$ in (\ref{dEE}) arises from the fact that the surface gravity $\kappa$ on apparent horizon is negative in the FRW universe, $i.e.$ $\kappa<0$ \cite{Akbar:2006kj}, which will impose a constraint on a perfect fluid in the framework of Einstein-Gauss-Bonnet gravity,  %obtained by
%\begin{eqnarray} \label{nkappaFRWa}
%1+\frac{p}{\rho}<\frac{4}{n-1}\Big(1+\frac{\widetilde\alpha}{R_A^2+\widetilde\alpha}\Big)\,,
%\end{eqnarray}
see Appendix \ref{A} for related discussions.

The entropy $S$ for the $n$-dimensional FRW universe in Einstein-Gauss-Bonnet gravity is given by
\citep{Cai:2005ra,Akbar:2006kj}\footnote{Entropy of stationary black holes can be derived using methods such as the Euclidean or Noether charge approach \cite{Iyer:1995kg}. Wald proposed a local geometric approach for dynamical black holes \cite{Wald:1993nt}, which Hayward further adapted using the Kodama vector \cite{Hayward:1998ee}. Although this approach could potentially be adapted to the FRW universe within the context of this modified gravity, it is notably more intricate and will therefore not be pursued in this paper.}
\begin{equation}\label{SEGB}
    S=\frac{A}{4G_{n}}\left[1+\frac{(n-2)}{(n-4)}\frac{2\widetilde{\alpha}}{R_A^{2}}\right]\,.
\end{equation}

In the same way as we did for Einstein gravity, here again we compare the Eq.(\ref{dEE}) with the standard form of thermodynamic first law \cite{Hayward:1997jp,Wu:2009wp}
\begin{alignat}{1}
d U=Td S-Pd V\,, \label{SFL1}
\end{alignat}
we see that the internal energy $U$ is identified with $-E$, $i.e.$ $U\equiv -E$, and the thermodynamic pressure $P$ with $W$, $i.e.$
\begin{eqnarray}\label{PW}
P\equiv&W\,.
\end{eqnarray}

After determining the definition of thermodynamic pressure, using Eqs.(\ref{HawkingT}), (\ref{WDa}) and (\ref{PW}), we further obtain the equation of state for the FRW universe in Einstein-Gauss-Bonnet gravity, $i.e.$\footnote{If $\widetilde{\alpha}<0$, the equation of state for $R_A=\sqrt{-2\widetilde{\alpha}}$ is independent of temperature, refering to a thermodynamic singularity. This behaviour is the same as in the modified gravity with a generalized conformal scalar field \cite{Kong:2021qiu,Abdusattar:2023pck,Abdusattar:2024alq}.
If the Gauss-Bonnet term is absent or  $\widetilde{\alpha}=0$, it simplifies to the result obtained in Einstein gravity (\ref{ES}). In the following discussion of this section, we will set $G_n = 1$ for simplicity.}
\begin{eqnarray}\label{PRTEGB}
P=\frac{(n-2)}{4} \Big(1+\frac{2\widetilde\alpha}{R^2_A}\Big)\frac{T}{R_A}+\frac{(n-2)(n-3)}{16\pi}\frac{1}{R^2_A}+\frac{(n-2)(n-5)}{16\pi}\frac{\widetilde\alpha}{R^4_A}\,.
\end{eqnarray}
% which coincides with the result discussed in \cite{Saavedra:2023lds}.
Solving the Eq.(\ref{PRTEGB}) with the critical condition (\ref{PVTc}), one can see that the appearance of $P$-$V$ phase transition is depended upon the spacetime dimension $n$ and gravitational coupling parameter $\widetilde\alpha$, see more related discussions in \cite{Saavedra:2023lds}.

As in the previous section, using Eqs. (\ref{SEGB}) and (\ref{PRTEGB}), we can calculate the heat capacity of the $n$-dimensional Gauss-Bonnet FRW universe
\begin{eqnarray}\label{CP}
C_{P}=T\left(\frac{\partial S}{\partial T}\right)_{P}=
\frac{(n-2)\pi^{(1+n)/2}T R_A^{n-3}(R_A^2 +2\widetilde\alpha)^2}{\Gamma[(n-1)/2][R_A^2 (n-3+2\pi R_A T)+2\widetilde\alpha(n-5+6\pi R_A T)]}\,.
\end{eqnarray}
We see that the sign of the denominator determines the stability of the universe because the numerator is always positive. A positive denominator implies a stable universe with positive heat capacity $C_P>0$, while a negative denominator implies an unstable universe with negative $C_P<0$. The expression of the heat capacity (\ref{CP}) can be also used to conveniently calculate the Joule-Thomson coefficient as shown in the following
section.

\section{Joule-Thomson Expansion of Arbitrary Dimensional FRW Universe}\label{secIV}

In this section, we study the Joule-Thomson expansion as an application of the thermodynamic equations of state to elucidate the cooling-heating properties of the $n$-dimensional FRW universe. Furthermore, we determine the inversion temperature and inversion pressure for the FRW universe, and illustrate the characteristics of inversion curves and isenthalpic curves in the $T$-$P$ plane.

\subsection{Joule-Thomson coefficient}

To study cooling/heating properties of the thermodynamic system, the Joule-Thomson coefficient $\mu$ is introduced, and is defined by the change of the temperature with respect to pressure under a fixed enthalpy given by \cite{Okcu:2016tgt,Feng:2020swq}
\begin{equation}\label{JTC0}
\mu=\large \left(\frac{\partial T}{\partial P}\large \right)_{\cal H} \,,
\end{equation}
where the sign of $\mu$ is determined by cooling or heating occurred in the thermodynamic system.
When $\mu > 0$ it signifies that the system is in a cooling process during the expansion while $\mu < 0$ the heating process occurs. Consequently, the points at which this cooling-heating transition occurs are defined by $\mu=0$, and at that specific temperature, we call it the inversion temperature, denoted as $T_i$. When the temperature of the system matches exactly with $T_i$, the corresponding pressure becomes the inversion pressure, denoted as $P_i$. This unique combination of temperature and pressure is known as the inversion point, denoted as ($T_i$, $P_i$).
The (\ref{JTC0}) can also be rewritten as following form
\begin{equation}\label{JTC}
\mu=\frac{1}{{C}_{P}}\left[T\left(\frac{\partial V}{\partial T}\right)_{P}-V \right] \,.
\end{equation}
%with $C_{P}$ being the specific heat capacity at constant pressure, which is an interesting quantity related to thermal stability of the thermodynamic system.
%There is a well-known property that the positive heat capacity
%corresponds to a stable system whereas a negative heat capacity indicates the instability of
%the system.

Use the Eqs.(\ref{EE}), (\ref{PRTEGB}) and (\ref{TV}), we obtain the enthalpy of $n$-dimensional FRW universe is given by\footnote{It can be directly deduced that the condition ${\cal H} < 0$ is equivalent to $\rho + p > 0$. This condition is consistent with the Null Energy Condition (NEC)- a fundamental physical constraint in cosmology, which is generally satisfied by most matter fields.}
\begin{eqnarray}\label{HH}
 {\cal H}&\equiv& U+P V \nonumber\\ %&=&-\frac{(n-2)\pi^{(n-3)/2}}{8\Gamma[(n-1)/2]}R_A^{n-3}\Big(1+\frac{\alpha}{R_A^2}\Big)+\frac{2\pi^{(n-1)/2}}{(n-1)\Gamma[(n-1)/2]}P R_A^{n-1}\\
 &=& \frac{(n-2)\pi^{(n-3)/2} R_A^{n-5} (R_A^2 +2 \widetilde\alpha) (2\pi R_A T-1)}{4(n-1) \Gamma[(n-1)/2]} \,,
\end{eqnarray}
see the related discussion about negativity/positivity condition of enthalpy in Appendix \ref{appB}.
From (\ref{HH}), one writes the temperature as a function of ${\cal H}$ and $R_A$ as follows
\begin{equation} \label{THRA}
T({\cal H},R_A)= \frac{1}{2\pi R_A} + \frac{2(n-1)\Gamma[(n-1)/2]\pi^{(1-n)/2}}{n-2} \frac{{\cal H}R_A^{4-n}}{R_A^2 +2\widetilde\alpha} \,,
\end{equation}
and then substituting the Eq.(\ref{THRA}) into (\ref{PRTEGB}), the pressure $P$ can be rewritten as a function of ${\cal H}$ and $R_A$,
\begin{equation} \label{PHRA}
P({\cal H},R_A)=(n-1)\left[\frac{\pi^{(1-n)/2}}{2}\Gamma\Big[\frac{n-1}{2}\Big]{\cal H}R_A^{1-n}+\frac{(n-2)}{16\pi R_A^2}\Big(1+\frac{\widetilde\alpha}{R_A^2}\Big)\right]  \,.
\end{equation}
By using (\ref{JTC0}) combined with Eqs.(\ref{THRA}) and (\ref{PHRA}), we obtain the Joule-Thomson coefficient of the $n$-dimensional FRW universe in the framework of Einstein-Gauss-Bonnet gravity
\begin{eqnarray}
\mu&=&\large \frac{\left({\partial T}/{\partial R_A}\large \right)_{{\cal H}}}{\large \left({\partial P}/{\partial R_A}\large \right)_{{\cal H}}}\nonumber\\
&=& \frac{4(n-2)\pi^{n/2}R_A^{n+3}(R_A^2+2\widetilde\alpha)^2 +16(n-1)\Gamma[(n-1)/2]\pi^{3/2}{\cal H}R_A^8 [(n-2)R_A^2 +2 (n-4)\widetilde\alpha]}{(n-2)(n-1)(R_A^2 +2\widetilde\alpha)^2 [4(n-1)\Gamma[(n-1)/2]\pi^{3/2}{\cal H}R_A^5 +(n-2)\pi^{n/2}R_A^n (R_A^2 +2\widetilde\alpha)]} \label{muH}\\
&=& \frac{2 R_A^2 \{R_A^2 [(n-2) (2 \pi R_A T-1)+1]+2 \widetilde\alpha[(n-4)(2 \pi  R_A T-1)+1]\}}{\pi T(n-1) (n-2) (R_A^2+2 \widetilde\alpha)^2}\,,\label{JTE1}
%
%\mu=\large \frac{\left({\partial T}/{\partial R_A}\large \right)_{{\cal H}}}{\large \left({\partial P}/{\partial R_A}\large \right)_{{\cal H}}}= \frac{8\Big[\frac{1}{2\pi R_A^2}+\frac{\xi {\cal H}R_A^{3-n}[(n-2)R_A^2 +2(n-4)\alpha]}{(n-2)(R_A^2 +\alpha)}\Big]}{(n-1)\Big[2\xi{\cal H}R_A^{-n} +\frac{(n-2)}{\pi R_A^3}\Big(1+\frac{\alpha}{R_A^2}\Big) \Big]} \,,
\end{eqnarray}
which can also be obtain by using Eqs.(\ref{PRTEGB}), (\ref{CP}) and (\ref{JTC}).
In particular, the Joule-Thomson coefficient diverges at $R_A=\sqrt{-2\widetilde{\alpha}}$ for $\widetilde{\alpha}<0$ which is consistent with the thermodynamic singularity as discussed in (\ref{PRTEGB}).

\subsection{Inversion temperature and inversion pressure}

Except for the singular point, the Joule-Thomson coefficient $\mu$ of $n$-dimensional FRW universe can be either positive or negative, implying that there exist both cooling and heating stages. In the following, we discuss the inversion temperature and inversion pressure for the $n$-dimensional FRW universe both in Einstein gravity and Einstein-Gauss-Bonnet gravity, respectively.

\begin{figure}[h]
\begin{minipage}[t]{7cm}
%\centering
\includegraphics[width=8cm,height=6cm]{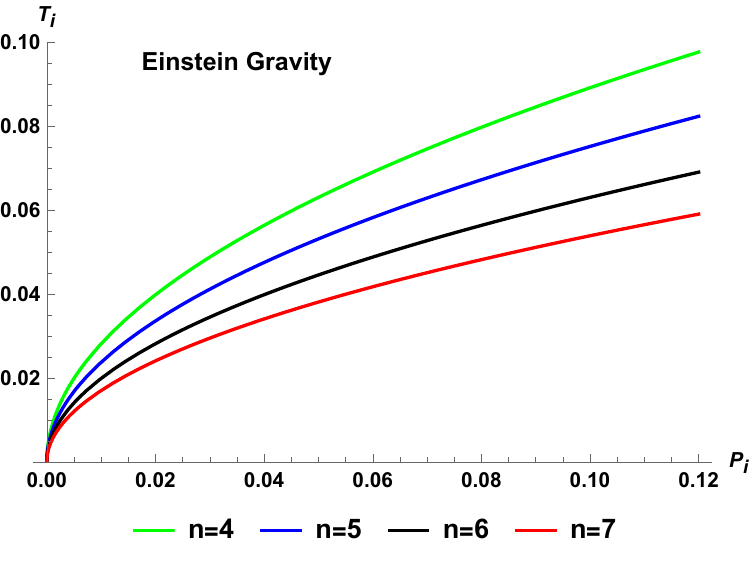}
\end{minipage}
 \caption{\label{FigInvTPa0} \footnotesize {\bf Einstein gravity:} The inversion curves of the FRW universe with dimensions $n=4,5,6,7$.}
\end{figure}

In the absence of the Gauss-Bonnet term with $\alpha=0$ (or $\widetilde{\alpha}=0$), we can derive the Joule-Thomson coefficient $\mu$ for an $n$-dimensional FRW universe in Einstein gravity from equations (\ref{muH}) and (\ref{JTE1}).
In this scenario, a positive enthalpy (${\cal H}>0$) leads to $\mu>0$, indicating that the FRW universe in arbitrary dimensions is always cooling, without inversion temperature or pressure.
By setting $\mu = 0$, we obtain the enthalpy for the existence of an inversion point given by
\begin{equation}\label{muH0}
{\cal H}=-\frac{\pi^{(n-3)/2}R_A^{n-3}}{4\Gamma[(n-1)/2] (n-1)}<0\,.
\end{equation}
Thus, from Eq.(\ref{JTE1}) with condition $\mu = 0$, we obtain the inversion temperature
\begin{equation}\label{invTa0}
T_{i}=\frac{n-3}{2(n-2)\pi R_A} \,
\end{equation}
and inversion pressure
\begin{equation}\label{invPa0}
P_{i}=\frac{n (n-3) }{16\pi R_A^2} \,
\end{equation}
of the $n$-dimensional FRW universe in Einstein gravity for the negative enthalpy, $i.e.$ ${\cal H}<0$.
From Eqs.(\ref{invTa0}) and (\ref{invPa0}), we also obtain the simple relation between the inversion temperature and the inversion pressure
\begin{equation}\label{invTP}
T_{i}=\frac{2\sqrt{n-3}}{n-2}\sqrt{\frac{P_i}{n\pi}} \,.
\end{equation}
Concern with the special one with $n=4$, we also obtain the result of four dimensional FRW universe in Einstein gravity \cite{Abdusattar:2021wfv}.
The inversion temperature $T_i$ of the FRW universe versus inversion pressure $P_i$ in Einstein gravity with  dimensions $n=4,5,6,7$ is shown in Fig.\ref{FigInvTPa0}.

We observe from Fig.\ref{FigInvTPa0} that, the inversion temperature rises as the inversion pressure increases. However, this rise becomes less slowly as the spacetime dimension $n$ increases, indicating a decreasing rate of change in the inversion temperature with respect to the inversion pressure.

In the presence of Gauss-Bonnet term with $\widetilde\alpha\neq 0$, and setting $\mu = 0$,
%from Eq.(\ref{JTE1}) we obtain the inversion temperature of FRW universe in Einstein-Gauss-Bonnet gravity from Eq.(\ref{JTE1}) as
from Eq.(\ref{muH}) we obtain
\begin{equation}\label{muH0}
    {\cal H}=-\frac{(n-2) \pi^{(n-3)/2} R_A^{n-5} (R_A^2+2\widetilde\alpha)^2}{4 \Gamma[(n-1)/2] (n-1) [(n-2)R_A^2 +2 (n-4)\widetilde\alpha]}\,.
\end{equation}
The enthalpy ${\cal H}<0$ for $\widetilde\alpha>0$, and relate this with the result given in (\ref{B6}) one can confirm that there is an inversion point. Additionally, by relating Eq.(\ref{muH0}) to (\ref{AppB7}) and (\ref{AppB10}), we find that the enthalpy ${\cal H}$ is also negative when $R_A>\sqrt{-2(n-4)\widetilde\alpha/{(n-2)}}$ for $\widetilde\alpha <0$. Conversely, when
$\widetilde\alpha <0$ and $R_A <\sqrt{-2(n-4)\widetilde\alpha/{(n-2)}}$, the enthalpy ${\cal H}$ is positive. However, given that the energy density and entropy given in Eqs.(\ref{rho1}), (\ref{SEGB}) are both inherently positive, no valid solution for $R_A$ arises in this case, implying that no inversion point can exist under these conditions.

\begin{figure}[h]
%\centering
\subfigure[]{
\begin{minipage}[t]{7cm}
%\centering
\includegraphics[width=7cm,height=5cm]{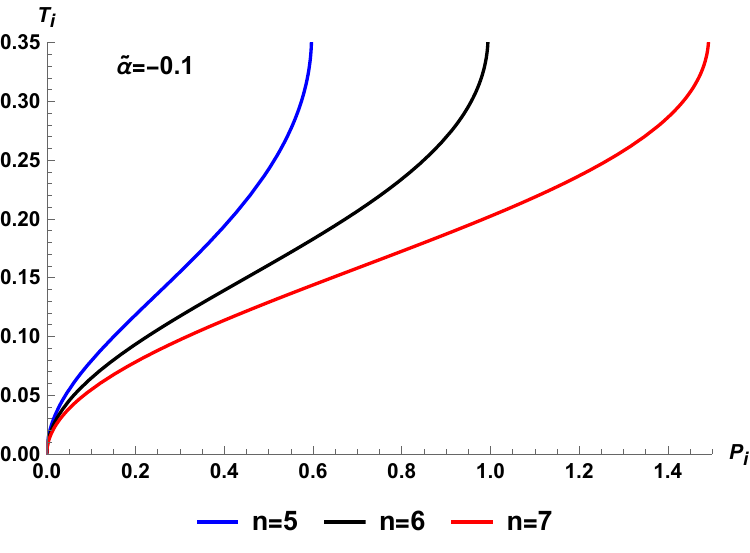}
\end{minipage}
}
\subfigure[]{
\begin{minipage}[t]{7cm}
%\centering
\includegraphics[width=7cm,height=5cm]{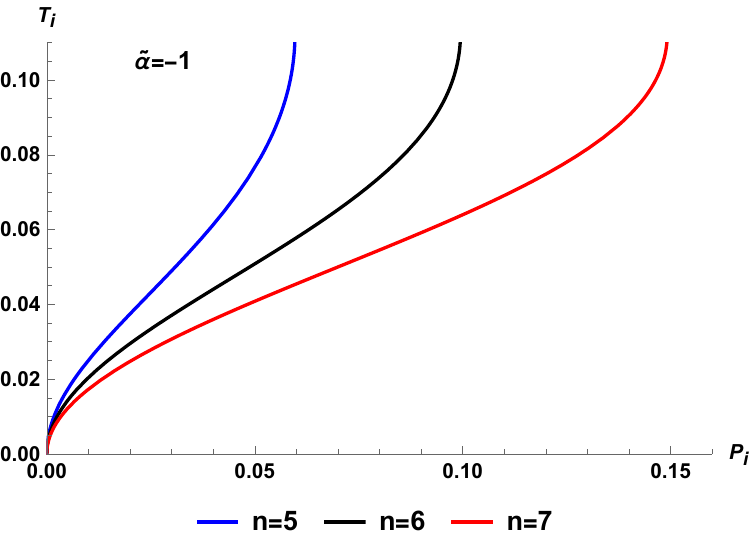}
\end{minipage}
}
\subfigure[]{
\begin{minipage}[t]{7cm}
%\centering
\includegraphics[width=7cm,height=5cm]{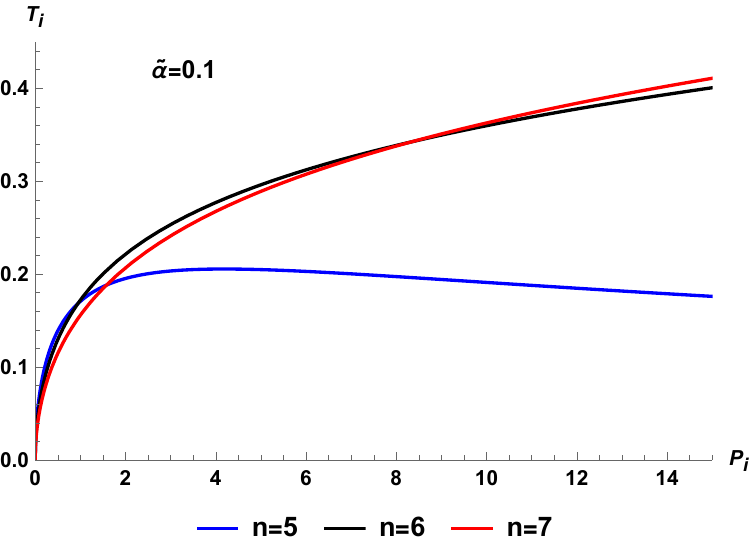}
\end{minipage}
}
\subfigure[]{
\begin{minipage}[t]{7cm}
%\centering
\includegraphics[width=7cm,height=5cm]{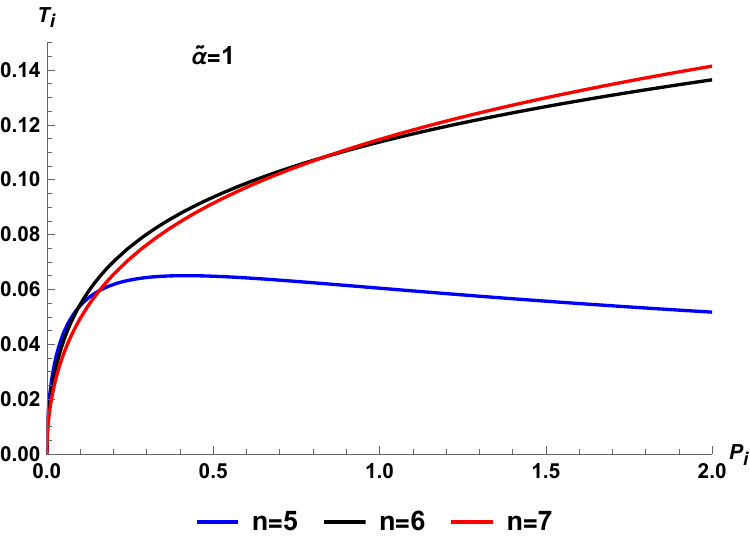}
\end{minipage}
}
 \caption{\label{FigInvTP} \footnotesize {\bf Einstein-Gauss-Bonnet gravity:} The inversion curves of the FRW universe with dimensions $n=5,6,7$.
 Figures (a) and (b) represent the cases where $\widetilde\alpha<0$, while figures (c) and (d) show the cases where $\widetilde\alpha>0$.}
\end{figure}

Setting $\mu = 0$, we also obtain the inversion temperature from Eq.(\ref{JTE1}) as
\begin{equation}\label{invT}
T_{i}=\frac{(n-3)(R_A^2+2\widetilde\alpha)-4\widetilde\alpha}{2\pi R_A[(n-2)R_A^2 +2(n-4)\widetilde\alpha]} \,.
\end{equation}
Substituting the expression (\ref{PRTEGB}) into (\ref{JTE1}), we can also get the inversion pressure given as
\begin{equation}\label{invP}
P_{i}=\frac{(n-2) \{n (n-3) R_A^4 +\widetilde\alpha [n (3 n-13)+2] R_A^2+
2n(n-5) \widetilde\alpha^2 \}}{16\pi R_A^4 [(n-2)R_A^2+2 (n-4) \widetilde\alpha]} \,.
\end{equation}
The inversion temperature $T_i$ of the FRW universe versus inversion pressure $P_i$ in Einstein-Gauss-Bonnet gravity with dimensions $n=5,6,7$ is shown in Fig.\ref{FigInvTP}.

We can observe from Fig.\ref{FigInvTP} that the inversion temperature of the FRW universe exhibits distinct behaviors depending on whether $\widetilde\alpha<0$ or $\widetilde\alpha>0$ as the inversion pressure varies. Specifically, as shown in Fig.\ref{FigInvTP}(a) and Fig.\ref{FigInvTP}(b), the inversion temperature increases as the inversion pressure rises for $\widetilde\alpha<0$. This increase becomes less slowly as the spacetime dimension increases, indicating a decreasing rate of change. In addition, the inversion temperature also increases as the value of the coupling parameter rises.
On the other hand, as shown in Fig.\ref{FigInvTP}(c) and Fig.\ref{FigInvTP}(d), the behavior of the inversion temperature is influenced by the spacetime dimension for $\widetilde\alpha>0$. In this case, the inversion temperature decreases as the coupling parameter increases, contrasting with the behavior observed for $\widetilde\alpha<0$, where the coupling parameter exerted an opposing influence on the inversion temperature.

\subsection{Isenthalpic curves}

The Joule-Thomson expansion is an isenthalpic process. Therefore, it is interesting to study the isenthalpic curves of the FRW universe. In the followings, we plot the inversion and isenthalpic curves of the FRW universe in $T$-$P$ plane with fixed ${\cal H}$ and $\widetilde{\alpha}$, and determine the cooling-heating region.

\begin{figure}[h]
%\centering
 \begin{minipage}[h]{1\linewidth}
\includegraphics[width=8cm]{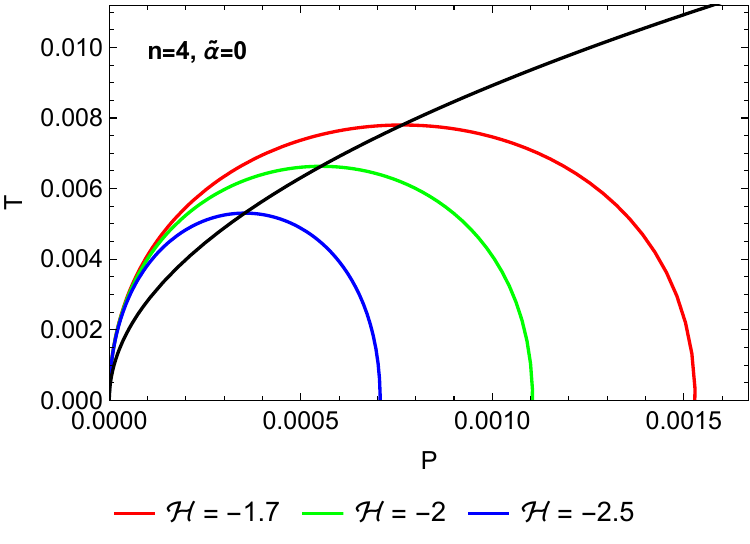}
 \put(-128,-10){(a)}
 \includegraphics[width=8cm]{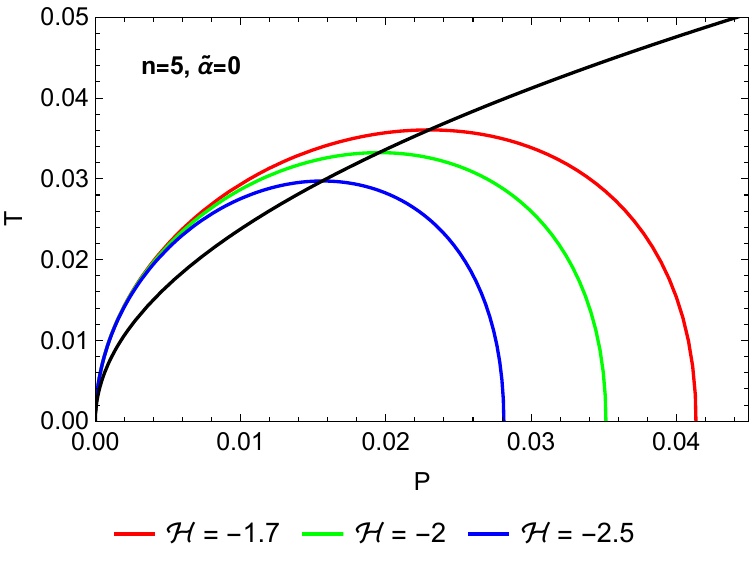}
 \put(-128,-10){(b)}
\end{minipage}
%\centering
 \begin{minipage}[h]{1\linewidth}
\includegraphics[width=8cm]{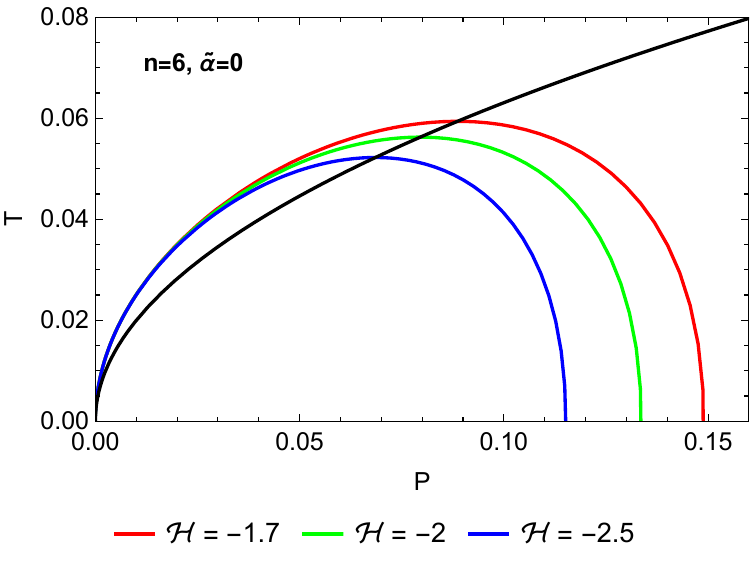}
\put(-128,-10){(c)}
\includegraphics[width=8cm]{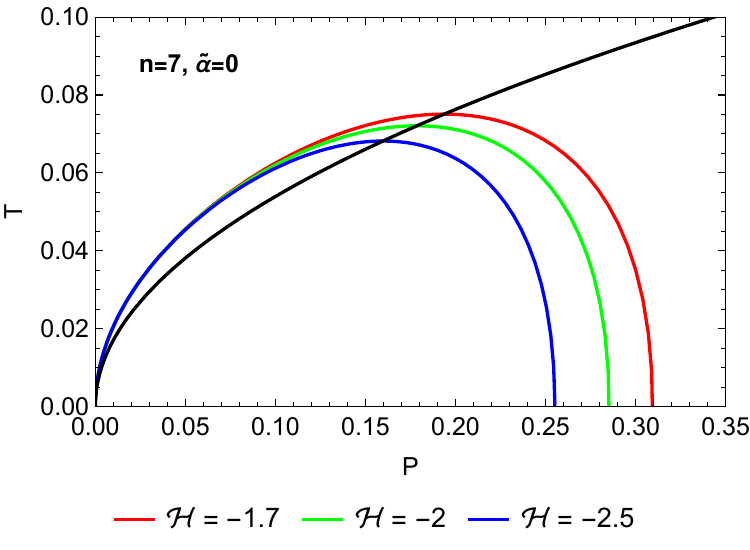}
\put(-128,-10){(d)}
\end{minipage}
 \caption{\label{FigJTa0} \footnotesize {\bf Einstein gravity:} The inversion and isenthalpic curves of the FRW universe with dimensions $n=4,5,6,7$.
The black line represents the inversion curve, and the rainbow curves intersect with it at points where they exhibit maxima. The values of ${\cal H}$ are marked on $T$-$P$ plane.}
\end{figure}
Figure \ref{FigJTa0} shows the inversion and isenthalpic curves of the FRW universe in Einstein gravity with $\widetilde{\alpha}=0$, which are plotted by Eqs.(\ref{THRA}), (\ref{PHRA}), (\ref{invTa0}) and (\ref{invPa0}).
Each graphics has three isenthalpic curves for
different values of enthalpy, the red solid curve, green solid curve and blue solid curve are ${\cal H}=-1.7, -2, -2.5$, respectively. The black solid curve represents the inversion curve, which is consistent with that in Fig.\ref{FigInvTPa0}. The inversion curves intersect with the peak of isenthalpic curves.
Naturally, it means that the left side of the isenthalpic curve exhibits a positive slope, whereas the slope turns negative on the right side. Consequently, during the throttling process, the inversion curve is regarded as the dividing line between the cooling and heating  regions. Additionally, through analysis and comparison, it becomes evident that the intersection point of the inversion curve and the isenthalpic curve in the FRW universe exhibits an upward trend as both enthalpy and spacetime dimension increase.

\begin{figure*}[ht]
%\centering
 \begin{minipage}[t]{1\linewidth}
\includegraphics[width=8cm]{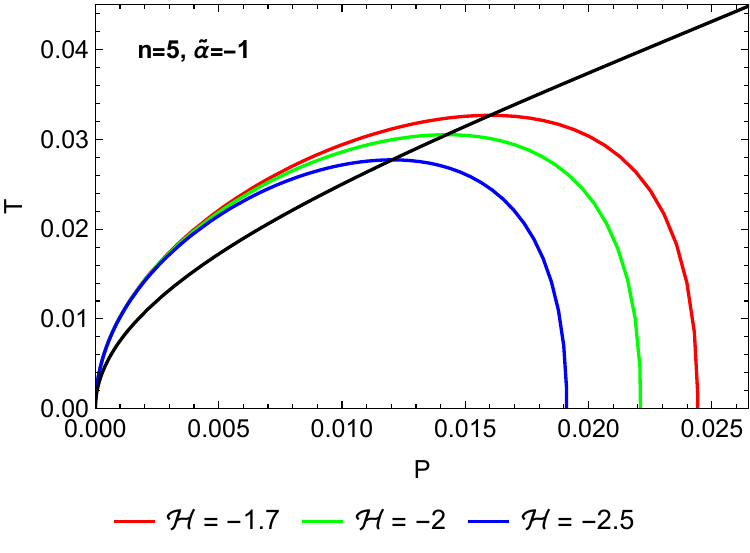}
 \put(-120,-10){(a)}
 \includegraphics[width=8cm]{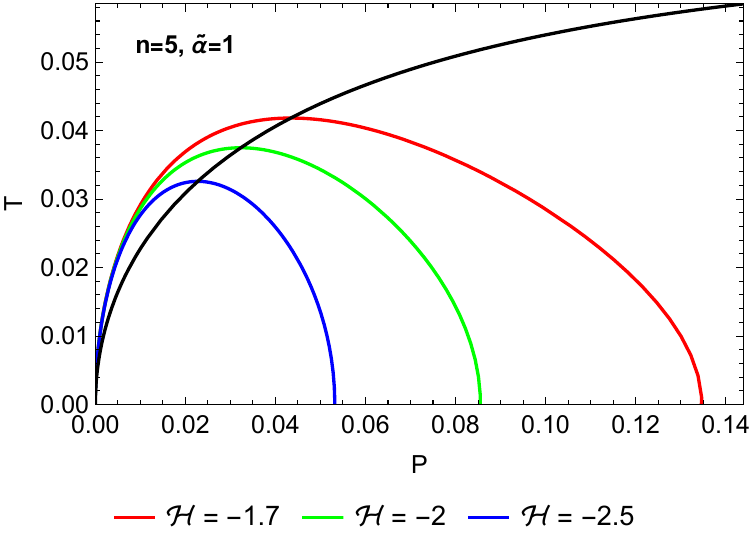}
 \put(-120,-10){(b)}
\end{minipage}
%\centering
 \begin{minipage}[t]{1\linewidth}
\includegraphics[width=8cm]{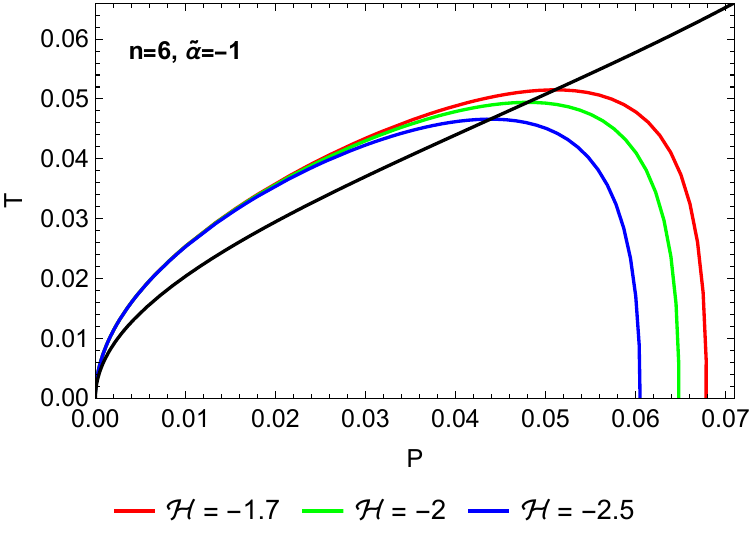}
\put(-120,-10){(c)}
\includegraphics[width=8cm]{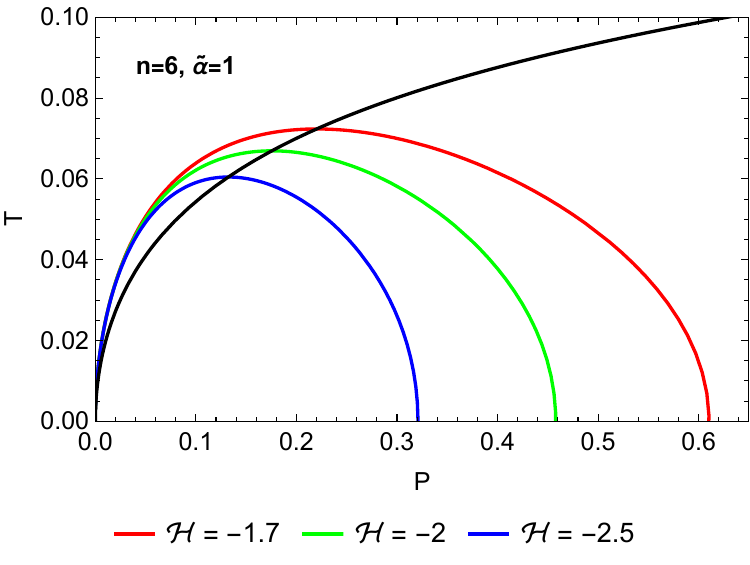}
\put(-120,-10){(d)}
\end{minipage}
%\centering
 \begin{minipage}[t]{1\linewidth}
\includegraphics[width=8cm]{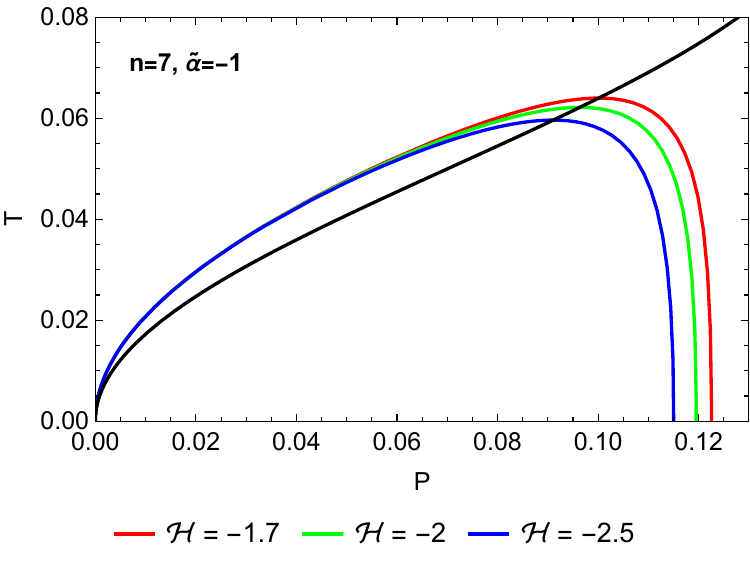}
\put(-120,-10){(e)}
\includegraphics[width=8cm]{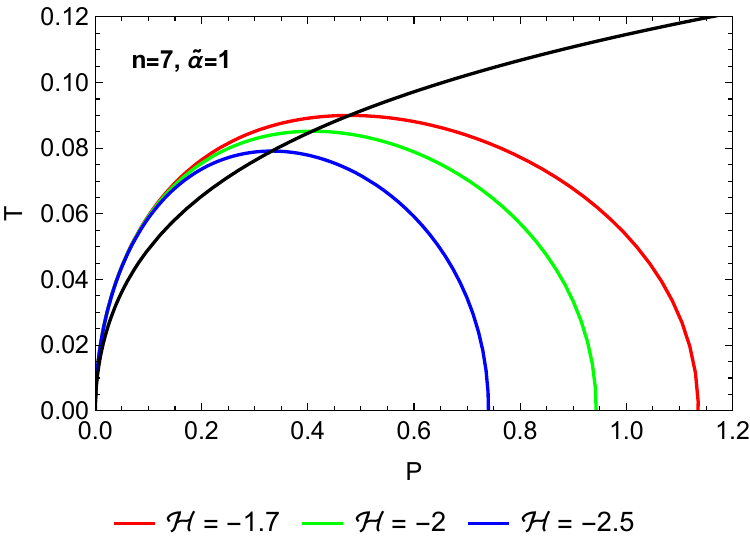}
\put(-120,-10){(f)}
\end{minipage}
 \caption{\label{FigJT} \footnotesize {\bf Einstein-Gauss-Bonnet gravity:} The inversion and isenthalpic curves of the FRW universe with dimensions $n=5,6,7$. The black line represents the inversion curve, and the rainbow curves intersect with it at points where they exhibit maxima. The values of ${\cal H}$ and $\widetilde
 {\alpha}$ are marked on $T$-$P$ plane.}
\end{figure*}

Figure \ref{FigJT} shows the inversion and isenthalpic curves of the FRW universe within the framework of Einstein-Gauss-Bonnet gravity, which are plotted by Eqs.(\ref{THRA}), (\ref{PHRA}), (\ref{invT}), (\ref{invP}).
It can be seen from Fig.\ref{FigJT} that under Einstein-Gauss-Bonnet gravity, the inversion point of the FRW universe increases with the enthalpy and spacetime dimension.
Through careful analysis, we also find that when the value of the gravitational coupling parameter approaches $0$, the behaviour of isenthalpic curves of the FRW universe is similar to that of Einstein's gravity. When the values of the coupling parameter deviate from $0$, the behavior of the isenthalpic curves of the FRW universe is obviously different from the Einstein gravity case. It is evident that the maximum value of the isenthalpic curve of the FRW universe increases with both enthalpy and spcatime dimension.

%%%%%%%%%%%%%%%%%%%%%%%%%%%%%
%The Joule-Thomson expansion occurs as an isenthalpic process, and hence it is significant to study the isenthalpic curves for a thermodynamic system. From Eqs.(\ref{PHRA}), (\ref{THRA}) and (\ref{invTP}), we will plot the isenthalpic curves in the Fig~\ref{FigJT} for a FRW universe in the $T$-$P$ plane by fixing the ${\cal H}$, and also show the inversion curve.
%\begin{figure}[h]
%\centering
% \begin{minipage}[t]{1\linewidth}
%\includegraphics[width=8cm]{MTP1.pdf}
%\end{minipage}
% \caption{\label{FigJT} \footnotesize The isenthalpic curves for a FRW universe have been plotted with different values of ${\cal H}$, while the monotonous black line is the inversion curve. These isenthalpic curves are separated by the inversion curve into two regions, i.e. the left cooling and right heating regions.}
%\end{figure}
%
%Obviously, these isenthalpic curves intersect with the inversion curve at their maximum points. In addition, these isenthalpic curves are also separated into two regions by the inversion curve, i.e. the left cooling and right heating regions. Furthermore, the region surrounded by isenthalpic curve and $P$ axis shrinks when ${\cal H}$ decreases or becomes more negative.

%%%%%%%%%%%%%%%%%%%
\subsection{Constraints on the Perfect Fluid in the FRW Universe from the Inversion Point ($\mu=0$)}

In what follows, we derive constraints on the perfect fluid of the FRW universe from the inversion point ($\mu=0$).

In the case where $\mathcal{H}<0$, the existence of an inversion point--corresponding to a peak in the $T$-$P$ diagram during Joule-Thomson expansion-is particularly noteworthy. Investigating the deeper physical implications of this phenomenon is crucial, and a simplified analysis is presented below.

By equating the Hawking temperature Eq.(\ref{HawkingT}) with the inversion temperature Eq.(\ref{invT}), we obtain
\begin{equation}
\frac{\dot{R}_A}{HR_A}=\frac{2(R_A^2 +2\widetilde\alpha)}{(n-2)R_A^2 +2(n-4)\widetilde\alpha}\,,
\end{equation}
which together with (\ref{dRA}) results to the following relation
\begin{equation}
\dot{H}-\frac{k}{a^2}=-\frac{2(R_A^2 +2\widetilde\alpha)}{R_A^2[(n-2)R_A^2 +2(n-4)\widetilde\alpha]}\,.
\end{equation}
Furthermore, by utilizing Eqs.(\ref{FE11}) and (\ref{FE10}), we ultimately derive the constraints on the perfect fluid in the FRW universe within the framework of Einstein-Gauss-Bonnet gravity as follows
\begin{equation}
1+\frac{p}{\rho}=\frac{4(R_A^2 +2\widetilde\alpha)^2}{(n-1)(R_A^2 +\widetilde\alpha)[(n-2)R_A^2 +2(n-4)\widetilde\alpha]}\,. \label{C2}
\end{equation}

When $\alpha=0$ (or $\widetilde{\alpha}=0$ (i.e., in the absence of the Gauss-Bonnet term), we also derive the constraints on the perfect fluid in the FRW universe within the framework of Einstein gravity as follows
%\footnote{It is also consistent with the result of constraint on the perfect fluid from negative surface gravity obtained in (\ref{AnkappaFRW}).}
\begin{equation}
1+\frac{p}{\rho}=\frac{4}{(n-1)(n-2)} \,. \label{C3}
\end{equation}
Specifically, when $n=4$, the expression (\ref{C3}) reduces to $\ddot{a}=0$ at the inversion point, indicating a transition between acceleration ($\ddot{a} > 0$) and deceleration ($\ddot{a} < 0$), or vice versa \cite{Abdusattar:2021wfv}.
These observations highlights the noticeable influence of spacetime dimensions on the perfect fluid and the evolution of the FRW universe.

\section{Conclusions and Discussions}\label{secV}

In this paper, we have studied the thermodynamics especially the equation of state and Joule-Thomson expansion of the $n$-dimensional FRW universe with a perfect fluid.
From the negative surface gravity of the FRW universe on apparent horizon, $i.e.$ $\kappa < 0$, we imposed a constraint on a perfect fluid in the framework of Einstein gravity and Einstein-Gauss-Bonnet gravity respectively, as detailed in Appendix \ref{A}.
We have derived the equations of state for the $n$-dimensional FRW universe, where the thermodynamic pressure $P$ is defined by the work density $W$, which is a natural definition directly read out from the first law of thermodynamics.
We find that the $n$-dimensional FRW universe under Einstein's gravity with perfect fluid do not have $P$-$V$ phase transitions and critical behavior. Moreover, we find that the specific capacity of the $n$-dimensional FRW universe is always positive. This result is remarkable, since it implies that the FRW universe is stable. Under Einstein-Gauss-Bonnet gravity, the appearance of $P$-$V$ phase transitions as well as (un)stability of the FRW universe with a perfect fluid are depended on the spacetime dimension $n$ and gravitational coupling parameter $\widetilde{\alpha}$.
%It is obvious that the modification term in gravitational field influence on the evolution of the universe.

Furthermore, we present the Joule-Thomson expansion as an application of thermodynamic equations of state to explain the thermodynamic properties of the $n$-dimensional FRW universe.
We derived the Joule-Thomson coefficient, and find that the equation of state from Einstein gravity shows that there is an inversion point for the FRW universe with arbitrary dimensions if its enthalpy ${\cal H}$ is negative. If the enthalpy ${\cal H}$ is positive (${\cal H}>0$), the FRW universe with arbitrary dimensions is always in the cooling stage, $i.e.$ there is no inversion temperature and inversion pressure.
As an extension, we also calculate the inversion temperature and inversion pressure for FRW universe in Einstein-Gauss-Bonnet gravity with arbitrary dimensions, and illustrate the characteristics of inversion curves and isenthalpic curves in the $T$-$P$ plane.
We observe that an inversion point exists for negative enthalpy (${\cal H}<0$), and its occurrence depends on both the spacetime dimension $n$ and the gravitational coupling parameter $\widetilde{\alpha}$. Moreover, we also find that the characteristics of inversion temperature exhibits distinct behaviors depending on whether $\widetilde\alpha<0$ or $\widetilde\alpha>0$ as the inversion pressure varies. Moreover, we examined the constraints on the perfect fluid of the FRW universe, derived from the cooling-heating transition point.
%, which align with the constraints stemming from negative surface gravity, as discussed in Appendix \ref{A}.

Notably, in the present work, we introduce the $T$-$P$ isenthalpic expansion process into a high-dimensional FRW universe within the framework of modified gravity, marking the first pioneering exploration of thermodynamic properties across arbitrary dimensions. This innovation not only establishes a theoretical foundation to support future astronomical observations--potentially providing empirical anchors for our findings-but also deepens our understanding of cosmic evolution from a thermodynamic perspective. Furthermore, a key subsequent challenge, yet one of significant import, lies in determining whether cosmological observations can detect an inversion temperature, a task that bridges our theoretical framework with observational cosmology and could further validate the thermodynamic insights we have uncovered. It would be interesting to extend the present study to higher-order modified gravity theories such as Lovelock gravity, as they may reveal intriguing phenomena. Another open question arises when considering scenarios where entropy may take on different values. In our current work, our primary focus is on the case where entropy is positive, and all our discussions and analyses are grounded on this premise. However, Refs. \cite{Clunan:2004tb,Cvetic:2001bk,Nojiri:2002qn} highlight that entropy can be negative due to the coupling of the Gauss-Bonnet term, presenting a fascinating avenue for future research to deepen our understanding of thermodynamic properties in Gauss-Bonnet gravity. It is therefore intriguing to explore the implications that would arise if we consider cases where entropy could be negative. We will pursue these investigations in the future.

\section{Acknowledgment}

We are deeply grateful to the reviewer for their insightful and constructive comments, which have significantly improved the quality of our work.
We also thank Profs.Ya-Peng Hu and Yu-Sen An for their useful discussions. Haximjan Abdusattar is supported by National Natural Science Foundation of China (NSFC) under grant No. 12465012, Kashi University high-level talent research start-up fund project under grant No. 022024002, and Tianchi Talented Young Doctors Program of Xinjiang Uygur Autonomous Region. Shi-Bei Kong is supported by grant No.12465011 from the National Natural Science Foundation of China (NSFC) and start-up fund DHBK2023002 from East China University of Technology (ECUT). Zhong-Wen Feng is supported by the Natural Science Foundation of Sichuan Province under Grant No. 24NSFSC1618.

%%%%%%%%%%%%%

\appendix

\section{Negativity Condition of Surface Gravity for the Arbitrary Dimensional FRW Universe} \label{A}

In this appendix, we derive the negativity conditions for the surface gravity (\ref{SurfaceG}) in both Einstein gravity and Einstein-Gauss-Bonnet gravity. By examining Eq.(\ref{SurfaceG}) or (\ref{HawkingT}), we deduce that a negative surface gravity ($\kappa|_{R=R_A}<0$) implies that the second term in these equations falls into one of the following categories: less than zero, exactly zero, or greater than zero but less than one. These conditions potentially impose constraints on the pressure and density characteristics of a perfect fluid.

Using Eqs.(\ref{RA}), (\ref{dRA}), and (\ref{FE11}), we obtain the second term in Eq.(\ref{SurfaceG}) as
\begin{equation}\label{AA1}
\delta=\frac{\dot{R}_A}{2H R_A}=-\frac{R_A^2}{2}\Big(\dot{H}-\frac{k}{a^2}\Big)=\frac{4\pi (\rho+p)}{n-2}\frac{R_A^2}{1+{2\widetilde\alpha}/{R_A^2}} \,.
\end{equation}

We analyze the following scenarios:

{\bf 1. $\delta=0$ ($\dot{R}_A=0$):}

This leads to $T=1/(2\pi R_A)$. Under this condition, we obtain $\rho+p=0$.

{\bf 2. $\delta<0$:}

This leads to $2\pi R_A T>1$. From Eq.(\ref{AA1}), we obtain
\begin{equation}\label{AA20}
\frac{4\pi (\rho+p)}{n-2}\frac{R_A^2}{1+{2\widetilde\alpha}/{R_A^2}}<0 \,.
\end{equation}
\begin{itemize}
  \item When $R_A < \sqrt{-2\widetilde\alpha}$, we have $\rho+p > 0$.
  \item When $\widetilde\alpha>0$, or $R_A > \sqrt{-2\widetilde\alpha}$ for $\widetilde\alpha<0$, we obtain $\rho+p<0$.
\end{itemize}

{\bf 3. $\delta=1$:}

This leads to $T=0$. From Eq.(\ref{AA1}), we obtain the corresponding negativity condition of surface gravity in Einstein-Gauss-Bonnet gravity
\begin{equation}\label{AA21}
\frac{4\pi (\rho+p)}{n-2}\frac{R_A^2}{1+{2\widetilde\alpha}/{R_A^2}}=1 \,,
\end{equation}
which is equivalent to
\begin{eqnarray} \label{AnkappaFRWa2}
\frac{p}{\rho}=1-\frac{4}{n-1}\Big(1+\frac{\widetilde\alpha}{R_A^2+\widetilde\alpha}\Big)\,,
\end{eqnarray}
where we have used the Eq.(\ref{rho1}).

{\bf 4. $0<\delta<1$:}

This leads to $0<2\pi R_A T<1$. From Eq.(\ref{AA1}), we obtain
\begin{equation}\label{AA22}
0<\frac{4\pi (\rho+p)}{n-2}\frac{R_A^2}{1+{2\widetilde\alpha}/{R_A^2}}<1 \,,
\end{equation}
\begin{itemize}
  \item When $R_A < \sqrt{-2\widetilde\alpha}$, we have
  \begin{eqnarray} \label{kappaA21}
\frac{n-2}{4\pi R_A^2}\Big(1+\frac{2\widetilde\alpha}{R_A^2}\Big)<\rho+p <0 \,,
\end{eqnarray}
which is equivalent to
\begin{eqnarray} \label{AnkappaFRWa1}
\frac{4}{n-1}\Big(1+\frac{\widetilde\alpha}{R_A^2+\widetilde\alpha}\Big)< 1+\frac{p}{\rho}<0\,.
\end{eqnarray}
  \item When $\widetilde\alpha>0$, or $R_A > \sqrt{-2\widetilde\alpha}$ for $\widetilde\alpha<0$, we obtain
  \begin{eqnarray} \label{kappaA22}
0<\rho+p<\frac{n-2}{4\pi R_A^2}\Big(1+\frac{2\widetilde\alpha}{R_A^2}\Big) \,,
\end{eqnarray}
which is equivalent to
\begin{eqnarray} \label{AnkappaFRWa2}
0<1+\frac{p}{\rho}<\frac{4}{n-1}\Big(1+\frac{\widetilde\alpha}{R_A^2+\widetilde\alpha}\Big)\,.
\end{eqnarray}
\end{itemize}

In the limit $\alpha\rightarrow 0$, the condition becomes
\begin{eqnarray} \label{AnkappaFRW}
-1<\frac{p}{\rho}<\frac{5-n}{n-1} \,,
\end{eqnarray}
which is the negativity condition for the surface gravity of the $n$-dimensional FRW universe in Einstein gravity.

\section{Negativity and Positivity of Enthalpy for the Arbitrary Dimensional FRW Universe}\label{appB}

Enthalpy is an important thermodynamic quantity that includes internal energy and the pressure-volume work to push surrounding systems. It has now become an effective tool for studying the thermodynamic behavior of gravitational systems, including the standard FRW spacetime.
This section will analyze the positivity and negativity of enthalpy in arbitrary dimensional FRW universes, with the core being to find the general conditions that determine its sign.

In Einstein gravity, using Eqs.(\ref{E}), (\ref{PV}) and (\ref{TV}), we obtain the enthalpy of the $n$-dimensional FRW universe as
\begin{eqnarray}\label{HHE}
 {\cal H}&\equiv& U+P V \nonumber\\
% &=&-\frac{(n-2)\pi^{(n-3)/2}}{8\Gamma[(n-1)/2]}R_A^{n-3}+\frac{2\pi^{(n-1)/2}}{(n-1)\Gamma[(n-1)/2]}P R_A^{n-1}\\
 &=& \frac{(n-2)\pi^{(n-3)/2} R_A^{n-3} (2\pi R_A T-1)}{4(n-1) \Gamma[(n-1)/2]} \,.
\end{eqnarray}
One can see that the enthalpy ${\cal H}>0$ when
\begin{equation}\label{HH1}
R_A T > \frac{1}{2\pi} \,,
\end{equation}
and the enthalpy ${\cal H}<0$ when
\begin{equation}\label{HH2}
 0<R_A T< \frac{1}{2\pi} \,.
\end{equation}

In Einstein-Gauss-Bonnet gravity, using Eqs.(\ref{EE}), (\ref{PRTEGB}) and (\ref{TV}), we obtain the enthalpy of the $n$-dimensional FRW universe
\begin{eqnarray}\label{HEGB}
 {\cal H}&\equiv& U+P V \nonumber\\ %&=&-\frac{(n-2)\pi^{(n-3)/2}}{8\Gamma[(n-1)/2]}R_A^{n-3}\Big(1+\frac{\alpha}{R_A^2}\Big)+\frac{2\pi^{(n-1)/2}}{(n-1)\Gamma[(n-1)/2]}P R_A^{n-1}\\
 &=& \frac{(n-2)\pi^{(n-3)/2} R_A^{n-5} (R_A^2 +2 \widetilde\alpha) (2\pi R_A T-1)}{4(n-1) \Gamma[(n-1)/2]} \,,
\end{eqnarray}
as given in Eq.(\ref{HH}).
\begin{description}
  \item[For $\widetilde\alpha>0$,]
the enthalpy ${\cal H}>0$ when
\begin{equation}
R_A T > \frac{1}{2\pi} \,,
\end{equation}
and the enthalpy ${\cal H}<0$ when
\begin{equation}\label{B6}
 0<R_A T< \frac{1}{2\pi} \,.
\end{equation}

  \item[For $\widetilde\alpha<0$,]
\end{description}
 \begin{itemize}
  \item the enthalpy ${\cal H}>0$ when
  \begin{equation}\label{AppB7}
   R_A>\sqrt{-2\widetilde\alpha}\,,
  \end{equation}
and negative ${\cal H}<0$
when
  \begin{equation}
   R_A<\sqrt{-2\widetilde\alpha}\,,
  \end{equation}
if $R_A T > {1}/{2\pi}$.
  \item the enthalpy ${\cal H}>0$ when
  \begin{equation}
   R_A<\sqrt{-2\widetilde\alpha}\,,
  \end{equation}
and negative ${\cal H}<0$
when
  \begin{equation}\label{AppB10}
   R_A>\sqrt{-2\widetilde\alpha}\,,
  \end{equation}
if $0<R_A T< {1}/{2\pi}$.
  \item the enthalpy ${\cal H}=0$ when
  \begin{equation}
   R_A=\sqrt{-2\widetilde\alpha}\,,
  \end{equation}
which corresponds to a thermodynamic singularity.
\end{itemize}

%%%%%%%%%%%%%%%%%%%%%

%%%%%%%%%%%%%%%%%%%%%%
%For $\tilde\alpha>0$, the temperature $T$ is positive and physical meaningful for $R_A<R_0$.

\end{document}